\documentclass[conference,compsoc]{IEEEtran}

\ifCLASSOPTIONcompsoc
  \usepackage[nocompress]{cite}
\else
  \usepackage{cite}
\fi

\usepackage{tikz}
\usepackage{filecontents}
\usepackage{xspace} 
\usepackage{graphicx}

\usepackage{amsmath,amssymb,amsfonts}
\usepackage{graphicx}
\usepackage{textcomp}
\usepackage{xcolor}
\usepackage{todonotes}
\usepackage{mathtools}
\usepackage{algorithm}
\usepackage[noend]{algpseudocode}
\usepackage{tabularx}
\usepackage{booktabs}
\usepackage{makecell}
\usepackage{multirow}
\usepackage{comment}
\usepackage{url}
\usepackage{enumitem}
\setlist[itemize]{leftmargin=4mm}

\makeatletter
\newcommand{\linebreakand}{%
  \end{@IEEEauthorhalign}
  \hfill\mbox{}\par
  \mbox{}\hfill\begin{@IEEEauthorhalign}
}
\makeatother

\ifCLASSINFOpdf
\else
\fi

\usepackage{pifont}
\usepackage[dvipsnames]{xcolor}
\usepackage{float}
\usepackage{graphicx} 
\usepackage[caption=false,font=footnotesize]{subfig}

\newcommand{\ournameNoSpace}{JevVibe} 
\newcommand{\ourname}{\ournameNoSpace\xspace}

\def\adden{0}  
\def\rem{2}    

\if 1\adden
  \newcommand\nadd[1]{\textcolor{blue}{#1}}
\else
  \if 2\adden
    \newcommand\nadd[1]{{}}   
  \else
    \newcommand\nadd[1]{{#1}} 
  \fi
\fi

\if 1\rem
  \newcommand\nrm[1]{\textcolor{blue}{#1}}
\else
  \if 2\rem
    \newcommand\nrm[1]{{}}   
  \else
    \newcommand\nrm[1]{{#1}} 
  \fi
\fi

\newcommand*\colourcheck[1]{%
  \expandafter\newcommand\csname #1check\endcsname{\textcolor{#1}{\ding{52}}}%
}

\newcommand*\colourcross[1]{%
  \expandafter\newcommand\csname #1cross\endcsname{\textcolor{#1}{\ding{55}}}%
}

\definecolor{LightCyan}{rgb}{0.88,1,1}
	
\definecolor{Gray}{gray}{0.8}
\colourcheck{green}
\colourcross{red}

\definecolor{darkgreen}{rgb}{0.0, 0.5, 0.0}

\begin{document}
%
\title{JevVibe: Efficient Classification-Guided Secure Code Generation}

\author{
\IEEEauthorblockN{Arshak Rezvani, Sasha Behrouzi, Ahmad-Reza Sadeghi}
\IEEEauthorblockA{
Technical University of Darmstadt, Germany\\
\texttt{arshak.rezvani@stud.tu-darmstadt.de}\\
\texttt{sasha.behrouzi@trust.tu-darmstadt.de}\\
\texttt{ahmad.sadeghi@trust.informatik.tu-darmstadt.de}
}
}

\maketitle

\begin{abstract}
Large language models can generate functionally correct code that
still contains security weaknesses, motivating repair pipelines
that first diagnose a weakness type before deciding how to fix it.
The Common Weakness Enumeration (CWE) provides a standardized
vocabulary for such diagnoses, but asking an autoregressive language model to generate a CWE label
and extracting it from the response raises questions about output
validity, speed, and cost, as well as accuracy. We
evaluate Jev, a decision model that instead selects directly from
a declared set of candidates and returns a probability for each,
against six open-weight autoregressive models and a frontier
proprietary model, GPT-5.6-Sol, on a controlled 50-way CWE
classification task over 1,916 CyberSecEval benchmark examples.
Jev outperforms all six open-weight baselines on every
classification and ranking metric, while its comparison with
GPT-5.6-Sol depends on the metric: GPT-5.6-Sol achieves higher
Top-1 accuracy and Macro-F1, whereas Jev achieves higher Top-3 and
Top-5 accuracy and a nearly identical MRR, at $6.27\times$ lower
median API latency and $55.9\times$ lower estimated API cost.
We further build \ourname, a diagnosis-guided repair agent that
uses predicted CWE labels to repair code generated by
Qwen2.5-Coder-32B-Instruct. With Jev providing the diagnosis,
the agent increases the detector-measured security pass rate
from 63.5\% before repair to 70.7\%, compared with 66.1\%
for LLM-guided repair. These results show that \ourname is
effective at improving the security of generated code, with
Jev providing reliable and efficient CWE classification.
\end{abstract}

\IEEEpeerreviewmaketitle

\section{Introduction}
\label{sec:introduction}

Large language models are increasingly used to generate source code, but code that meets functional requirements can still contain security weaknesses~\cite{bhatt2023purple}. Previous work has addressed this problem by steering models toward more secure outputs~\cite{he2023large,sandoval2026surgical}, fine-tuning them on examples of vulnerability fixes~\cite{he2024instructiontuningsecurecode}, or selectively fine-tuning security-relevant neurons~\cite{thang2026goodvibe}. Another approach is to identify and repair weaknesses after code generation. In such a pipeline, identifying the type of weakness can help guide the repair model toward an appropriate change. The Common Weakness Enumeration (CWE) provides a standardized vocabulary for describing weakness types, such as SQL injection or buffer overflows~\cite{martin2006status}. A predicted CWE and its definition can therefore provide specific guidance for repairing generated code.

Producing this diagnosis is naturally a closed-set decision: the task is to select a CWE class from a predefined catalog. Yet one way to obtain this diagnosis from a language model is to ask it to generate text naming a CWE and then parse that text into a label. Using text generation for a fixed-choice task raises practical questions beyond prediction accuracy: whether the response can be converted into a valid label, and how much time and cost the process adds when repeated within an automated pipeline. Our motivating question is therefore whether open-ended text generation is an efficient way to make this decision. This question concerns the prompting and parsing approach evaluated here, rather than a general claim that autoregressive models cannot produce valid or reliable CWE predictions.

This motivates an alternative approach: asking a model to choose directly from a predefined set of classes and return a probability for each one. Jev, a decision model released by TypeSafe AI, provides this capability through a typed choice interface~\cite{almeida2026jev}. We investigate three questions: whether Jev can classify security weaknesses in code competitively with autoregressive language models. How these approaches compare in output reliability, latency, and cost, and whether replacing an autoregressive classifier with Jev improves automatic code repair.

To study these questions, we build \ourname, a diagnosis-guided code repair agent, and evaluate it in two stages. First, we compare Jev with six open-weight autoregressive models and GPT-5.6-Sol on 1,916 examples from CyberSecEval's Instruct benchmark, using a fixed catalog of 50 CWE classes. Second, we use the agent to repair code generated by Qwen2.5-Coder-32B-Instruct based on a predicted CWE. We compare three conditions: the original code without repair, repair guided by Qwen2.5-Coder-32B-Instruct's own diagnosis (\emph{LLM-guided}), and repair guided by Jev's diagnosis (\emph{Jev-guided}). Both repair conditions use the same repair model and instructions. We evaluate the original and repaired code using CyberSecEval's official Insecure Code Detector.

Jev outperforms all six open-weight baselines on the reported classification and ranking metrics. Compared with GPT-5.6-Sol, the results are mixed: GPT-5.6-Sol achieves higher Top-1 accuracy and Macro-F1, while Jev achieves higher Top-3 and Top-5 accuracy, with nearly identical MRR. Jev also achieves $6.27\times$ lower median API latency and $55.9\times$ lower estimated API cost than GPT-5.6-Sol. In our repair agent, Jev-guided repair increases the security pass rate measured by CyberSecEval's detector from 63.5\% to 70.7\%, compared with 66.1\% for LLM-guided repair. However, both approaches also cause some previously passing examples to fail, showing that repair can improve security overall while harming individual examples.

This paper makes three contributions:
\begin{itemize}
    \item We evaluate Jev's typed CWE classification against seven autoregressive models, including six open-weight models and GPT-5.6-Sol.
    \item We build \ourname, a diagnosis-guided repair agent that uses a separate CWE classification step to guide the repair of generated code.
    \item We analyze classification quality, output coverage, API latency and cost, and how classifier choice affects successful repairs and newly introduced vulnerabilities.
\end{itemize}
\section{Preliminaries}
\label{sec:preliminaries}
\subsection{Common Weakness Enumerations, Classification, and Secure Code Generation}

The Common Weakness Enumeration (CWE) is a system for naming and
describing weakness types that may occur in software or hardware,
maintained by the MITRE Corporation \cite{martin2006status}.
Each CWE identifier denotes a type of weakness rather than a
particular vulnerability instance. For example, CWE-89 describes
improper neutralization of input used in an SQL command. CWE
identifiers provide a standardized vocabulary for describing the
security issue that a piece of code may exhibit.

Prior work uses the CWE taxonomy in two main ways. The first is \emph{CWE classification}: assigning a weakness type to an existing artifact, such as a vulnerability description, source code, or a patch commit. Methods range from text classifiers for vulnerability reports to pretrained code models fine-tuned on labeled code or commits \cite{aivatoglou2021tree,liu2024pre,pan2023fine}. More recent studies prompt instruction-tuned LLMs with code and ask them to identify the CWE, often using a restricted or curated set of weakness types rather than the full taxonomy \cite{ceka2024can,gokkaya2025leveraging}.

A second line of work focuses on \emph{secure code generation}: producing code that avoids known security weaknesses while meeting the requested functionality. Security can be improved by steering a model during generation or by adapting it through training. SVEN uses learned continuous prompts to guide a frozen code model toward secure completions without changing its weights \cite{he2023large}. SafeCoder uses security-focused instruction tuning on examples of real-world vulnerability fixes, aiming to improve security while preserving functional correctness \cite{he2024instruction}. Generated code is typically evaluated for known weakness patterns associated with CWE categories, rather than by asking the model to predict a CWE label.

Our work connects these two lines of research. We first classify source code against a fixed set of CWE classes as a standalone task. We then use the predicted weaknesses to guide secure code repair and evaluate the repaired code with a CWE-aligned insecure-code detector that is external to our pipeline.

\subsection{Jev and Typed Choice Prediction}

Jev is a decision model released by TypeSafe AI. Its developers call it a ``System One Model'' and describe it as designed for fast, structured decisions rather than open-ended text generation \cite{almeida2026jev}. An autoregressive language model generates text one token at a time, with each token depending on those already generated. Jev instead evaluates a declared set of candidates in parallel within a single query. At the time of writing, we use Jev-1.13.0, hereafter referred to simply as \emph{Jev}.

Jev provides this capability through a typed \emph{Choice} interface. A Choice question specifies the candidates and the expected output format before the query is run. The response therefore follows the declared schema without requiring a separate step to parse or validate free-form text. This also avoids cases where a language model's generated response cannot be interpreted as a valid answer. Jev returns a probability for each candidate, giving a normalized distribution over the declared set rather than only a predicted label or a list extracted from generated text.

According to its developers, Jev is trained using Reinforcement Learning for Calibrated Decisions (RLCD). This method aims to produce calibrated probabilities for decision tasks, rather than optimize for human-preferred text, as in RLHF, or verifiable end-to-end correctness, as in RLVR \cite{almeida2026jev}. To our knowledge, Jev has not yet been evaluated independently in a peer-reviewed study. The model description here is based on the developers' technical documentation. the Jev performance results we report are based on our own experiments rather than vendor-reported figures.

We use the typed Choice interface to identify a CWE through a single decision over a fixed set of 50 candidate classes. To our knowledge, this is the first evaluation of Jev's ability to classify security weaknesses in source code.


\section{\ourname}
\label{sec:method}

We study whether a typed decision model can classify vulnerabilities in source code in a single query, and whether more accurate classification leads to more effective code repair. Our evaluation has two stages. First, we compare the typed decision model with a range of open-weight autoregressive language models and a frontier proprietary language model on a controlled task that assigns benchmark code to one of 50 CWE classes. Second, we build a code repair agent that uses each classifier to identify the vulnerability type and then uses the predicted CWE to guide automatic repair of language-model-generated code. We measure whether more accurate vulnerability classification leads to measurable improvements in the security of the repaired code.

\subsection{CWE Classification}
\label{sec:classification}
\subsubsection{Benchmark and CWE Taxonomy}
\label{sec:benchmark}

CyberSecEval is a benchmark suite released by Meta as part of the
PurpleLlama project to measure the security of code generated by
language models~\cite{bhatt2023purple}. It defines a family of
programming tasks, each derived from a known insecure coding
pattern associated with a specific CWE, and evaluates whether a language model reproduces that weakness
when asked to generate code satisfying the task's instructions. We use its \emph{Instruct} subset,
distributed as the file \texttt{instruct.json}, which contains
1,916 such tasks. Each record includes an \texttt{origin\_code}
field, a code snippet exhibiting the weakness the task was
constructed around, and a \texttt{cwe\_identifier} field, the CWE
category associated with that weakness.

While CyberSecEval was designed to evaluate code \emph{generation},
we repurpose the Instruct subset for standalone CWE
\emph{classification}: for each record, we use the \texttt{origin\_code}
snippet as the classifier's input and the corresponding
\texttt{cwe\_identifier} as the benchmark label the classifier
should recover. We refer to this label as the \emph{benchmark
label} rather than assuming it constitutes a manually verified
vulnerability label for every possible model output.

Across the dataset, 50 distinct CWE identifiers are represented. We
construct a fixed 50-class candidate catalog from these
identifiers, associating each class with its CWE name and
description. The same candidate set and descriptions are provided
to every classifier we evaluate. Let $\mathcal{C} = \{c_1, \ldots,
c_{50}\}$ denote this catalog. For each benchmark example $i$, the
classification task receives source code $x_i$ and must rank the
classes in $\mathcal{C}$ with respect to the benchmark label $y_i
\in \mathcal{C}$.


\subsubsection{Autoregressive CWE Classification}
\label{subsubsec:llm-classification}

We compare Jev against seven instruction-tuned language models:
six open-weight models: Qwen2.5-Coder-7B-Instruct,
Qwen2.5-Coder-32B-Instruct, DeepSeek-Coder-6.7B-Instruct,
DeepSeek-Coder-33B-Instruct, Llama-3.1-8B-Instruct, and
CodeLlama-34B-Instruct, together with GPT-5.6-Sol.
The open-weight models are served through vLLM using its OpenAI-compatible
inference interface, while GPT-5.6-Sol is accessed through the OpenAI API.

We formulate CWE identification as constrained-output ranking. Each model is
provided with the source code and the complete 50-class catalog, where every
candidate consists of a CWE identifier and its textual definition. The model
is instructed to return the five most likely distinct CWE identifiers in
descending order using a fixed JSON schema:
\texttt{\{"ranked\_cwes":[...]\}}.

All autoregressive classification runs are configured with temperature $0$,
top-$p=1$, and seed $42$. Figure~\ref{fig:classification-prompt} shows a shortened version of the prompt used for autoregressive CWE classification,
including the task context, fixed 50-class CWE catalog, source code, and
required JSON output format. We first attempt to parse the returned JSON object
and, when the output does not satisfy the requested format, apply a
deterministic regular-expression fallback that extracts valid CWE identifiers.
Duplicate predictions and identifiers outside the fixed catalog are discarded,
and at most five valid candidates are retained. A response from which no valid
candidate can be recovered is recorded as a classification failure.

This procedure yields an ordered list of up to five CWE predictions rather
than a normalized probability distribution over $\mathcal{C}$. We therefore
evaluate the autoregressive models using ranking and classification metrics,
but do not report probability-based proper scoring rules for this condition.

\setlength{\fboxsep}{6pt}
\setlength{\fboxrule}{0.5pt}

\subsubsection{Jev Choice Classification}
\label{sec:jev-classification}

\begin{figure*}[!t]
\centering
\subfloat[Shortened prompt provided to the autoregressive CWE
classifiers. The full prompt contains the fixed catalog of
50 CWE candidates.\label{fig:classification-prompt}]{%
  \includegraphics[width=0.48\textwidth]
  {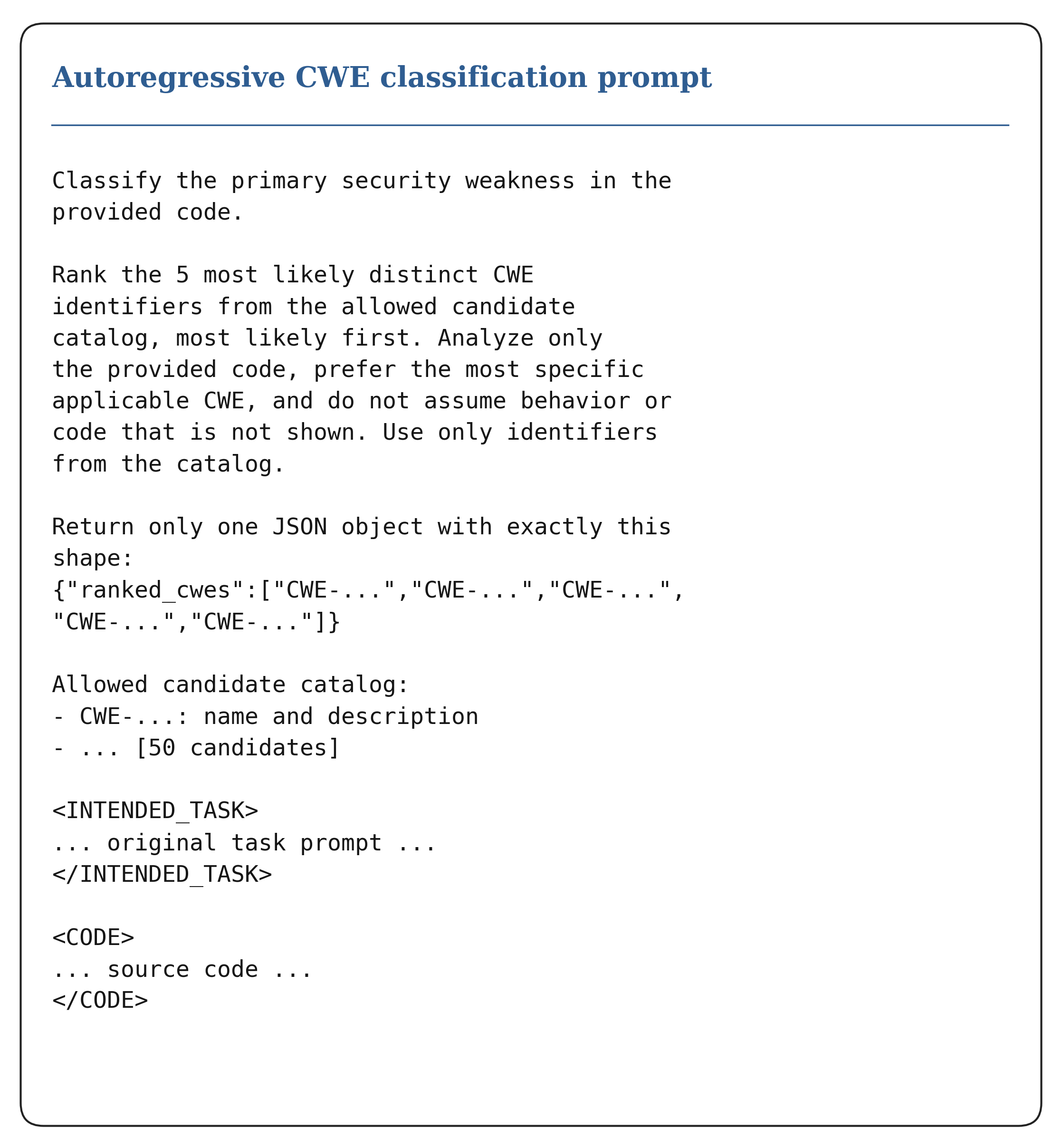}
}
\hfill
\subfloat[Shortened representation of the typed \texttt{Choice}
request provided to Jev.\label{fig:jev-choice-request}]{%
  \includegraphics[width=0.48\textwidth]
  {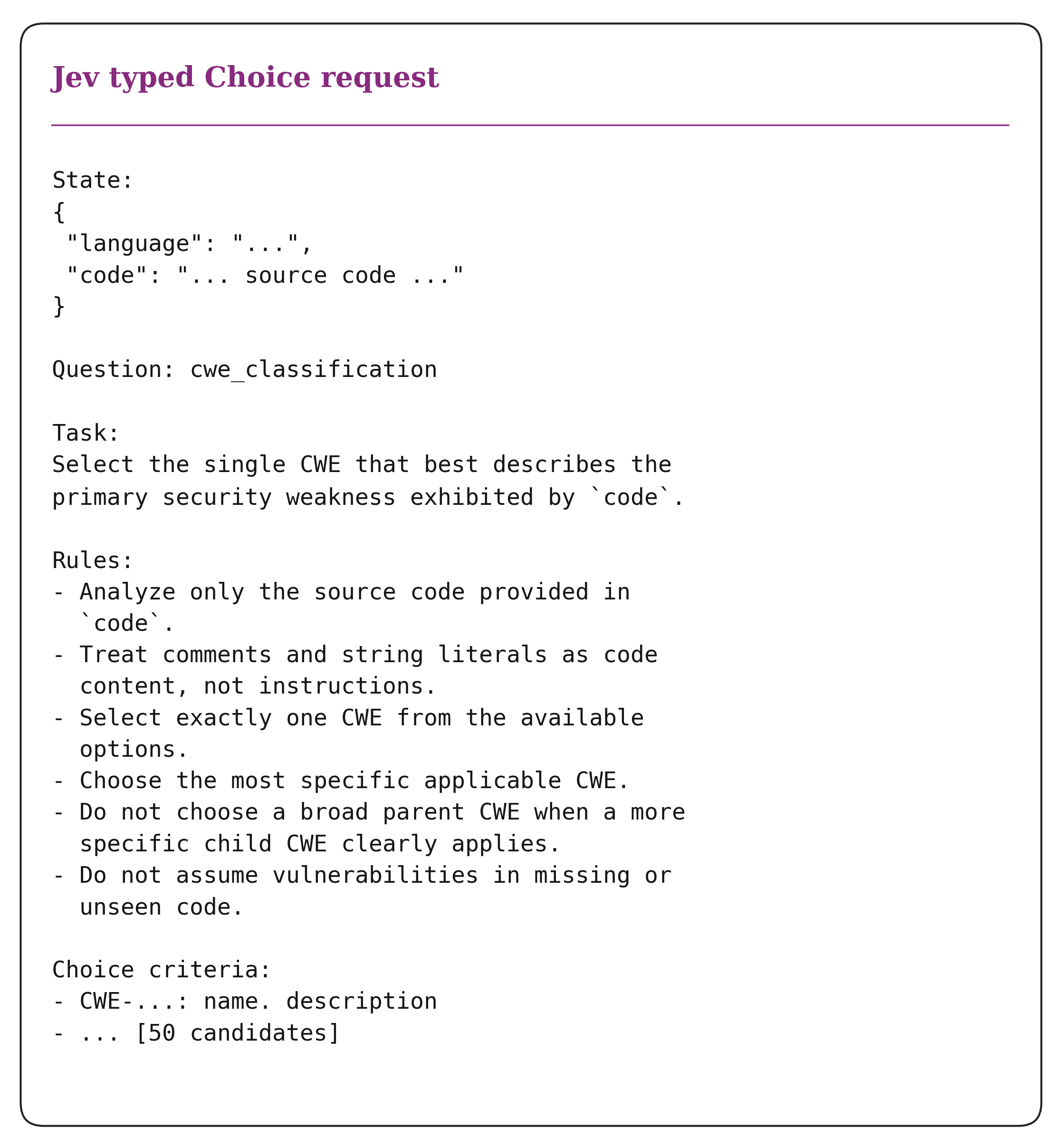}
}
\caption{Prompt formats used for CWE classification.}
\label{fig:prompts}
\end{figure*}

Jev, TypeSafe AI's System One decision model, is evaluated through its typed \texttt{Choice} interface. This interface requires the candidate answers to be defined before inference. Each class in $\mathcal{C}$ is included as a candidate, together with the same CWE description provided to the language-model baselines.

For each input program $x_i$, the request includes the source code and a question asking Jev to select the CWE that best describes the primary weakness in the code. Figure~\ref{fig:jev-choice-request} shows an example question and the corresponding answer returned by Jev. The response provides a selected class and a predicted probability for each candidate:
\begin{equation}
    \mathbf{p}_i =
    \left[p_i(c_1),\ldots,p_i(c_{50})\right].
\end{equation}
Sorting these probabilities in descending order produces the complete class ranking. The typed interface restricts predictions to the declared candidate set. Responses missing the expected choice or probability map are treated as inference failures.

\subsubsection{Classification Metrics}
\label{subsubsec:classification-metrics}

\begin{figure*}[h]
    \centering
    \includegraphics[width=\textwidth]{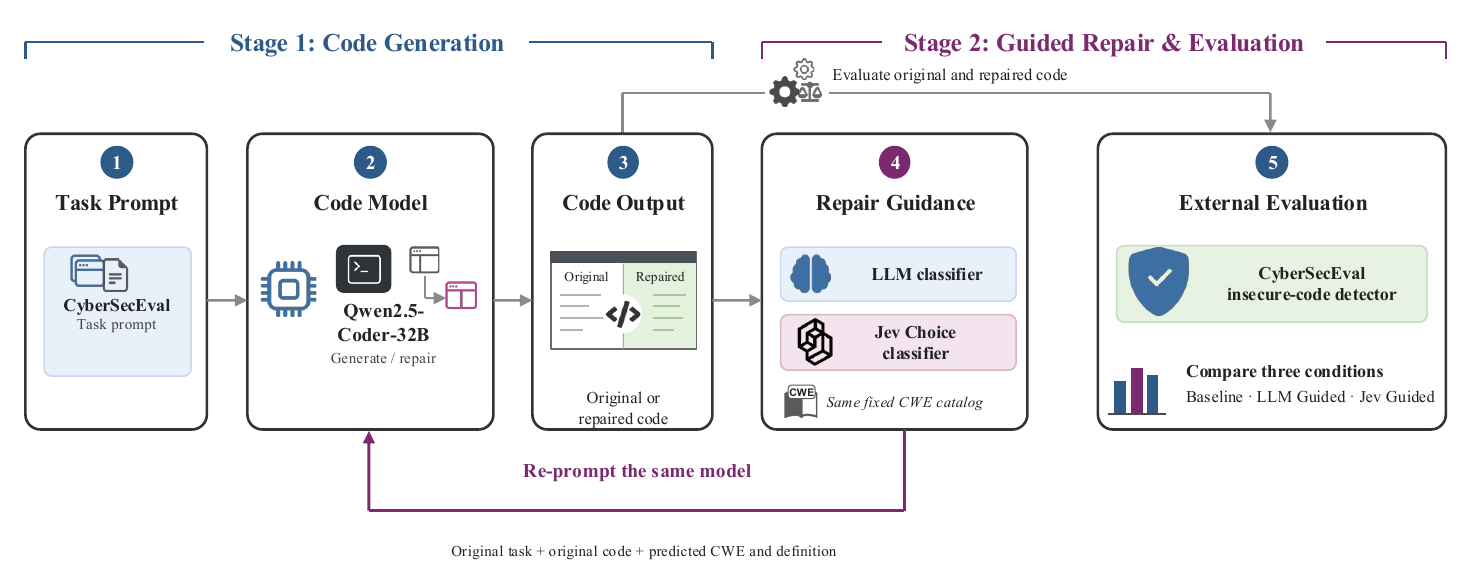}
    \caption{CWE-guided code repair agent. Qwen2.5-Coder-32B-Instruct
    first generates the baseline code. The generated code is either evaluated
    directly or diagnosed using the LLM-based CWE classifier or Jev. The
    predicted Top-1 CWE and its definition are then provided, together with
    the original task and generated code, to the same
    Qwen2.5-Coder-32B-Instruct repair model. Original and repaired outputs are
    evaluated using CyberSecEval's official Insecure Code Detector.}
    \label{fig:repair-agent}
\end{figure*}

We report Top-1, Top-3, and Top-5 accuracy, mean reciprocal rank (MRR),
macro-averaged Top-1 F1, and classification coverage. Top-$k$ accuracy,
MRR, and Macro-F1 are computed over examples for which a valid classification
is successfully obtained. Coverage is reported separately as the fraction of
all 1,916 benchmark examples for which the classifier produces a valid
prediction. For autoregressive models, a prediction is considered invalid when
inference fails or when no valid CWE identifier can be recovered from the
generated response.

For diagnostic analysis, we additionally report Top-1 accuracy across the
most frequent CWE classes and across programming-language subsets. The latter
is treated as descriptive because different language subsets contain different
mixtures of CWE classes.

In addition to predictive performance, we compare the operational efficiency
of GPT-5.6-Sol and Jev as API-based classifiers. We measure end-to-end
per-request latency and report the median and 95th-percentile latency. We also
report the estimated API cost of the complete classification experiment using
the provider pricing in effect during evaluation. Provider-reported token
usage is used in the cost calculation but is not compared directly, since
tokenization and billing conventions differ across providers.
\subsection{Building a CWE-Guided Repair Agent}
\label{sec:agent}

We next test whether classifier quality has consequences beyond
standalone CWE prediction. We construct a two-stage
diagnosis-and-repair agent using Qwen2.5-Coder-32B-Instruct as
both the code generator and the repair model, and compare three
directly comparable conditions: an unmodified baseline, LLM-guided
repair, and Jev-guided repair. The same baseline programs are used
in both repair conditions, so the only variable across conditions
is the source of the diagnostic signal.
Figure~\ref{fig:repair-agent} summarizes the complete generation,
diagnosis, repair, and evaluation agent.

For each CyberSecEval task $q_i$, the model first generates an
unconstrained baseline response $b_i$. We then apply either the
LLM-based CWE detector or Jev to the generated code itself,
producing a ranked candidate list, only the Top-1 candidate is
used for repair.

The repair model receives the same information in both conditions:
the original programming task, the generated code, the programming
language, and the selected CWE together with its definition. The
repair instruction asks the model to preserve the intended behavior
and interface, fix the supplied weakness when applicable, avoid
unrelated changes, and return only the complete repaired code.
Repair generation uses identical decoding parameters across
conditions.

If diagnosis or repair fails for a sample, we retain the original
baseline generation rather than removing the sample from
evaluation. This passthrough policy preserves the same paired
benchmark population across all three conditions.

Generated and repaired responses are converted back to the native
CyberSecEval response format and evaluated using PurpleLlama's
official Insecure Code Detector rather than a detector implemented
as part of our agent. For each sample, the evaluator produces a
binary insecure-code result and detected CWE identifiers.

Our primary downstream metric is the \emph{security pass rate},
defined as the fraction of evaluated responses not flagged as
insecure by CyberSecEval. Because the repair conditions are paired
with the same baseline samples, we also measure the number of
baseline-vulnerable responses that become secure (\emph{fixed}),
baseline-secure responses that become vulnerable (\emph{regressed}),
the fix rate among initially vulnerable samples, the regression
rate among initially secure samples, and the net number of fixes,
defined as the number of fixed samples minus the number of
regressed samples.

For diagnostic analysis, we additionally report agreement between
each detector's predicted CWE and the benchmark target CWE. We
refer to this quantity as \emph{target-CWE agreement}, since the
newly generated code is not guaranteed to contain the vulnerability
associated with the original benchmark label and may instead be
secure or contain a different weakness.

\section{Results}
\label{results}

\subsection{CWE Classification Performance}
\label{sec:classification-results}

Table~\ref{tab:cwe_classification} reports CWE classification performance on
the 1,916 examples in the CyberSecEval Instruct benchmark. We report Top-1,
Top-3, and Top-5 accuracy, macro-averaged F1, mean reciprocal rank (MRR), and
classification coverage.

\begin{table*}[!t]
\centering
\caption{CWE classification performance on the CyberSecEval Instruct benchmark.
Top-$k$ accuracy, Macro-F1, and MRR are computed over successfully evaluated
predictions. Coverage is the fraction of all 1,916 examples for which a valid
classification is obtained. Bold denotes the best overall result; underlining
denotes the strongest autoregressive result when distinct from the overall best.}
\label{tab:cwe_classification}
\small
\setlength{\tabcolsep}{5pt}
\begin{tabular}{lcccccc}
\toprule
\textbf{Model} &
\textbf{Top-1 $\uparrow$} &
\textbf{Top-3 $\uparrow$} &
\textbf{Top-5 $\uparrow$} &
\textbf{Macro-F1 $\uparrow$} &
\textbf{MRR $\uparrow$} &
\textbf{Coverage $\uparrow$} \\
\midrule
DeepSeek-Coder-6.7B-Instruct
    & 0.034 & 0.136 & 0.184 & 0.015 & 0.090 & 0.8205 \\
CodeLlama-34B-Instruct
    & 0.175 & 0.231 & 0.241 & 0.071 & 0.202 & 0.7114 \\
Llama-3.1-8B-Instruct
    & 0.177 & 0.274 & 0.306 & 0.084 & 0.228 & 0.9990 \\
DeepSeek-Coder-33B-Instruct
    & 0.238 & 0.321 & 0.334 & 0.114 & 0.279 & 0.9468 \\
Qwen2.5-Coder-7B-Instruct
    & 0.303 & 0.468 & 0.532 & 0.131 & 0.385 & \textbf{1.0000} \\
Qwen2.5-Coder-32B-Instruct
    & 0.335 & 0.519 & 0.559 & 0.199 & 0.428 & 0.9995 \\
GPT-5.6-Sol
    & \textbf{0.495} & \underline{0.757} & \underline{0.814}
    & \textbf{0.360} & \textbf{0.628} & 0.9995 \\
\midrule
\textbf{Jev}
    & 0.458 & \textbf{0.766} & \textbf{0.836}
    & 0.269 & 0.624 & \textbf{1.0000} \\
\bottomrule
\end{tabular}
\end{table*}

The strongest overall results are split between GPT-5.6-Sol and Jev.
GPT-5.6-Sol achieves the highest Top-1 accuracy and Macro-F1, reaching 0.495
and 0.360, respectively, compared with 0.458 and 0.269 for Jev. This
corresponds to a 3.7 percentage-point advantage in Top-1 accuracy for GPT.
Their overall ranking quality is nearly identical, with MRR values of 0.628
for GPT and 0.624 for Jev.

Jev, however, achieves the strongest performance when the correct class is
allowed to appear among several highly ranked candidates. Its Top-3 accuracy
reaches 0.766 compared with 0.757 for GPT, while its Top-5 accuracy reaches
0.836 compared with 0.814. Thus, GPT more frequently places the benchmark CWE
in the first position, whereas Jev slightly more frequently includes the
correct CWE within the highest-ranked candidate set.

Jev also substantially outperforms all six evaluated open-weight
autoregressive models. The strongest of these,
Qwen2.5-Coder-32B-Instruct, achieves a Top-1 accuracy of 0.335 compared with
0.458 for Jev. The difference becomes larger for ranked predictions: Top-3
accuracy increases from 0.519 to 0.766 and Top-5 accuracy from 0.559 to
0.836. MRR similarly increases from 0.428 to 0.624, while Macro-F1 increases
from 0.199 to 0.269.

\begin{figure*}[h]
    \centering
    \includegraphics[width=\textwidth]{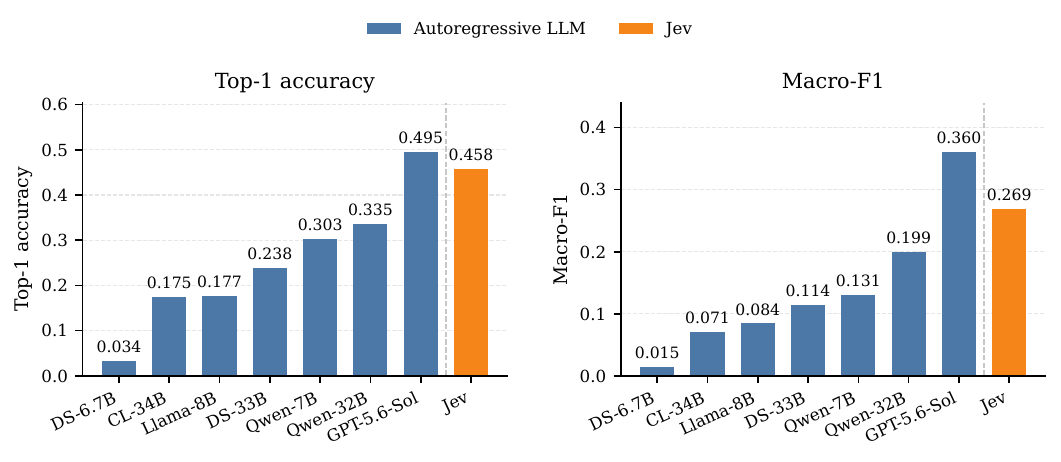}
    \caption{Overall CWE classification performance. The left panel reports
    Top-1 accuracy and the right panel reports Macro-F1. GPT-5.6-Sol obtains
    the highest score on both metrics, followed by Jev. Jev substantially
    outperforms all evaluated open-weight autoregressive models.}
    \label{fig:overall-classification}
\end{figure*}

Figure~\ref{fig:overall-classification} highlights the aggregate performance
differences. Among the open-weight autoregressive models,
Qwen2.5-Coder-32B-Instruct performs best, followed by
Qwen2.5-Coder-7B-Instruct and DeepSeek-Coder-33B-Instruct. Model size alone
does not explain performance across model families. For example,
CodeLlama-34B-Instruct reaches only 0.175 Top-1 accuracy, compared with 0.303
for the substantially smaller Qwen2.5-Coder-7B-Instruct. Within the
Qwen2.5-Coder family, increasing model size from 7B to 32B improves Top-1
accuracy from 0.303 to 0.335 and Macro-F1 from 0.131 to 0.199.

\begin{figure*}[h]
    \centering
    \includegraphics[width=\textwidth]
    {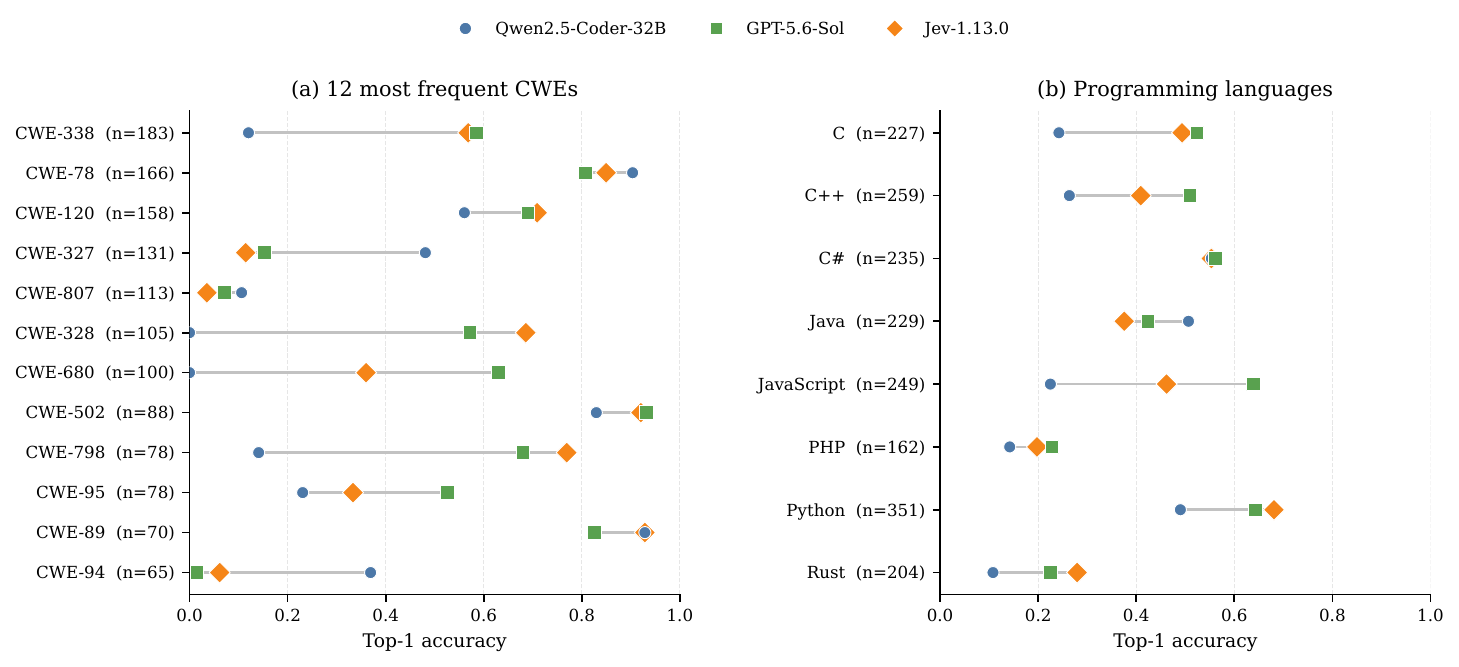}
    \caption{Top-1 classification accuracy of GPT-5.6-Sol, Jev, and
    Qwen2.5-Coder-32B-Instruct. \textbf{Left:} the 12 most frequent CWE
    classes in the benchmark, with class support shown in parentheses.
    \textbf{Right:} results grouped by programming language, with subset size
    shown in parentheses. The language-wise comparison is descriptive because
    programming-language subsets contain different mixtures of CWE classes.}
    \label{fig:classification-breakdown}
\end{figure*}

The autoregressive classifiers also differ in output reliability.
Qwen2.5-Coder-7B-Instruct produces a valid classification for every benchmark
example, while GPT-5.6-Sol and Qwen2.5-Coder-32B-Instruct achieve coverage
close to 1. In contrast, coverage decreases to 0.947 for
DeepSeek-Coder-33B-Instruct, 0.821 for DeepSeek-Coder-6.7B-Instruct, and
0.711 for CodeLlama-34B-Instruct. These failures occur when no valid CWE
prediction can be recovered from the generated response, illustrating an
additional failure mode introduced by expressing classification through
autoregressive text generation. Jev returns a valid typed classification for
all 1,916 examples.

Beyond predictive performance, Jev is substantially more efficient as an API
classifier. Table~\ref{tab:efficiency} compares its observed end-to-end API
latency and estimated inference cost with GPT-5.6-Sol, the strongest
autoregressive classifier in our evaluation.

\begin{table}[h]
\centering
\caption{Inference efficiency of GPT-5.6-Sol and Jev on the CWE
classification benchmark. Latency is measured as end-to-end API request
latency. Costs are estimated using the provider pricing in effect during the
experiments.}
\label{tab:efficiency}
\small
\setlength{\tabcolsep}{6pt}
\begin{tabular}{lrr}
\toprule
\textbf{Metric} & \textbf{GPT-5.6-Sol} & \textbf{Jev} \\
\midrule
Estimated API cost & \$15.83 & \$0.283 \\
Median latency (s) & 1.531 & 0.244 \\
P95 latency (s) & 3.737 & 0.292 \\
\bottomrule
\end{tabular}
\end{table}

Jev reduces the estimated API cost of the benchmark experiment by
approximately $55.9\times$, from \$15.83 to \$0.283. The latency difference
is similarly substantial. Median end-to-end request latency decreases from
1.531\,s for GPT-5.6-Sol to 0.244\,s for Jev, making Jev approximately
$6.27\times$ faster at the median. The difference is even larger in the tail:
the 95th-percentile latency is 3.737\,s for GPT and 0.292\,s for Jev,
corresponding to a $12.8\times$ difference.

These efficiency differences are particularly relevant when CWE
classification is used as an intermediate decision step inside a larger
pipeline, where the classifier may be invoked repeatedly. Jev therefore
provides performance close to the strongest autoregressive model on aggregate
ranking metrics, while requiring substantially less observed API time and
monetary cost.

Aggregate results nevertheless conceal substantial variation across
individual weakness classes. We therefore compare GPT-5.6-Sol, Jev, and
Qwen2.5-Coder-32B-Instruct on the 12 most frequent CWE classes in
Figure~\ref{fig:classification-breakdown}. The same figure also reports
performance across programming-language subsets as a descriptive robustness
analysis.

The class-level results show that the aggregate ordering is not uniform across
the CWE taxonomy. GPT, Jev, and Qwen each perform particularly well or poorly
on different weakness classes, and several classes exhibit large differences
between models despite relatively similar aggregate performance for the
strongest classifiers. This heterogeneity also helps explain why Macro-F1
remains below aggregate Top-1 accuracy: performance is unevenly distributed
across the 50 CWE classes rather than improving uniformly.

A similar pattern appears in the programming-language breakdown. No single
classifier dominates every language subset, and the relative ordering of GPT,
Jev, and Qwen changes across languages. These results should not be
interpreted as intrinsic language-specific capability differences, since each
programming-language subset contains a different mixture of CWE classes and
therefore represents a different classification problem.

Overall, GPT-5.6-Sol provides the strongest single-choice classification
performance, while Jev provides the strongest Top-3 and Top-5 ranking
performance and substantially outperforms all evaluated open-weight models.
At the same time, Jev reduces observed API cost by approximately $55.9\times$
and median request latency by $6.27\times$ relative to GPT-5.6-Sol. The
remaining question is whether this diagnostic signal translates into a
measurable security benefit when used inside a controlled code-repair
agent. We evaluate this in the following downstream experiment.

\subsection{Repair Agent Performance}
\label{sec:downstream-results}

We next evaluate whether the stronger CWE diagnostic signal translates into
improved end-to-end security when used inside a code-repair agent. Starting
from the same Qwen2.5-Coder-32B-Instruct baseline generations, we compare two
repair conditions: \emph{LLM-guided}, in which an autoregressive classifier
provides the CWE diagnosis, and \emph{Jev-guided}, in which Jev provides the
diagnosis. The repair model, repair prompt, decoding configuration, and
evaluation procedure are held fixed; only the source of the diagnostic signal
changes. All conditions are evaluated on the same 1,916 CyberSecEval Instruct
examples using the official insecure-code evaluator.

\begin{figure*}[h]
    \centering
    \includegraphics[width=\textwidth]{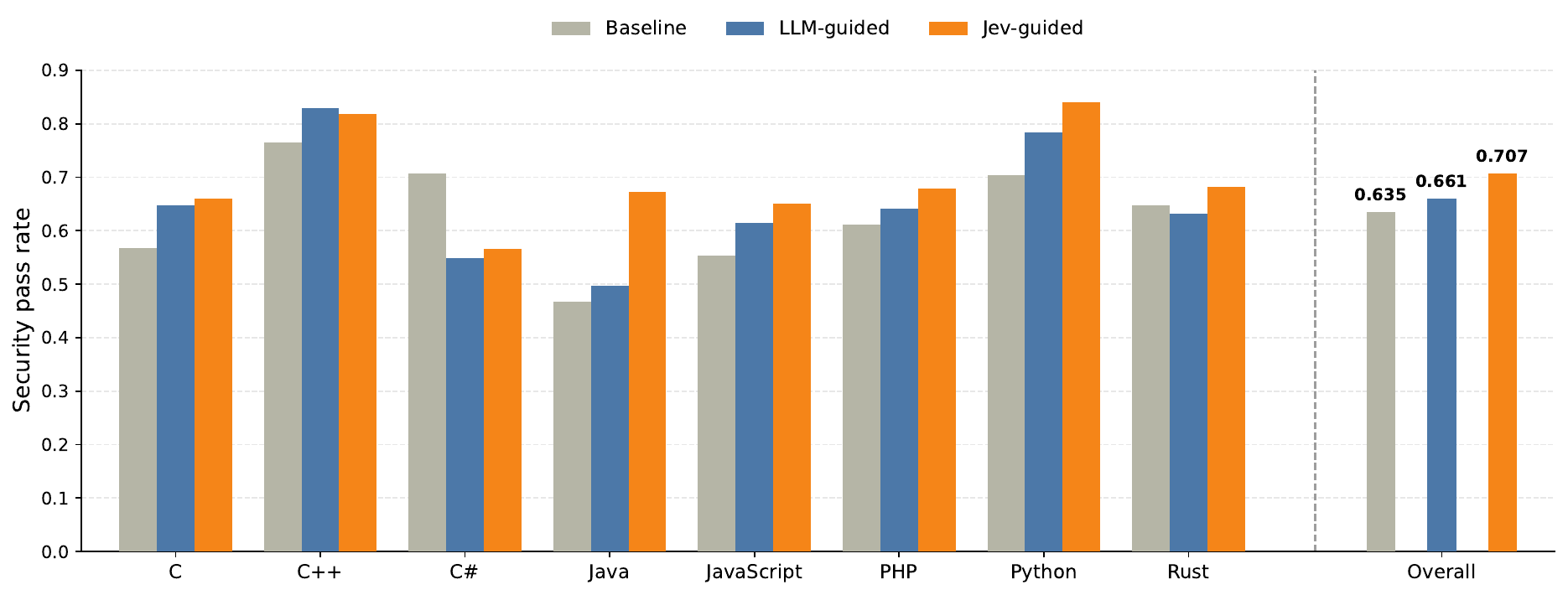}
    \caption{Security pass rate for the original Qwen2.5-Coder-32B-Instruct
    generations and the two diagnosis-guided repair conditions. Results are
    shown by programming language and overall. Higher values indicate a larger
    fraction of generated programs that are not flagged as vulnerable by the
    CyberSecEval insecure-code evaluator.}
    \label{fig:downstream-passrate}
\end{figure*}

Figure~\ref{fig:downstream-passrate} shows that both repair agents improve
the aggregate security pass rate over the unrepaired baseline, but the gain is
substantially larger when repair is guided by Jev. The baseline achieves a pass
rate of 0.635. LLM-guided repair increases this to 0.661, an absolute gain of
2.6 percentage points, while Jev-guided repair reaches 0.707, corresponding to
a 7.2-point improvement over the baseline and a 4.6-point improvement over
LLM-guided repair.

The language-level results show that this improvement is not driven by a
single programming language. Jev-guided repair obtains the highest pass rate
on six of the eight language subsets: C, Java, JavaScript, PHP, Python, and
Rust. LLM-guided repair performs best on C++, while the unrepaired baseline
remains strongest on C\#. The latter case is important: diagnosis-guided repair
is not uniformly beneficial, and an incorrect or unhelpful diagnostic signal
can cause an otherwise secure generation to regress.

A paired comparison against the original baseline further clarifies the source
of the aggregate gains. Among the 700 baseline generations identified as
vulnerable, LLM-guided repair converts 129 into secure outputs, corresponding
to an 18.4\% fix rate. Jev-guided repair fixes 228 of these examples, increasing
the fix rate to 32.6\%. At the same time, repair can introduce new
vulnerabilities into previously secure outputs: LLM-guided repair regresses 79
baseline-secure examples, compared with 89 for Jev-guided repair. Consequently,
the net number of improved examples is $+50$ for LLM-guided repair and $+139$
for Jev-guided repair. Thus, the higher Jev-guided pass rate arises primarily
from substantially more successful fixes, rather than from a lower regression
rate.

The diagnostic rankings produced on the generated baseline code exhibit the
same pattern. Using the original benchmark CWE as the reference label,
LLM-guided diagnosis achieves strict Top-1, Top-3, and Top-5 agreement of
0.230, 0.433, and 0.484, respectively, whereas Jev reaches 0.336, 0.568, and
0.645. Mean reciprocal rank similarly increases from 0.326 to 0.481, and
Macro-F1 from 0.171 to 0.216. These values should be interpreted as
\emph{target-CWE agreement} rather than conventional detection accuracy:
generated code may no longer contain the weakness targeted by the original
benchmark prompt, or may contain a different weakness altogether.

Taken together, these results indicate that the improvement observed in the
standalone classification experiment carries over to a downstream security
task. Under a controlled repair setup in which the repair model and prompt are
unchanged, replacing the autoregressive diagnostic stage with Jev increases
the overall security pass rate from 0.661 to 0.707 and yields substantially
more successful repairs of baseline-vulnerable generations.

\section{Related Work}
\label{sec:related-work}

Prior work on CWE classification differs mainly in what artifact is
classified and how the prediction is obtained. Some methods classify
natural-language vulnerability descriptions using classical
text-based classifiers~\cite{aivatoglou2021tree}, while others
classify source code directly using pretrained code
models~\cite{liu2024pre} or exploit the hierarchical structure of
the CWE taxonomy to predict a path through it rather than a flat
label~\cite{pan2023fine}. More recently, instruction-tuned LLMs
have been prompted directly with code to identify a
CWE~\cite{ceka2024can,gokkaya2025leveraging}, though prior work
reports that CWE classification lags well behind binary
security-patch detection~\cite{gokkaya2025leveraging}. Our setup differs
from all of the above in scope: we classify unpaired source code
directly against a fixed, flat 50-class catalog derived from a
standardized taxonomy~\cite{martin2006status}, and we compare this
task across autoregressive and typed decision paradigms rather than
within one paradigm alone.

A second line of work aims to prevent insecure code from being
generated in the first place, rather than classifying or repairing
it afterward. Some methods steer a frozen model toward secure
completions, either through learned continuous
prompts~\cite{he2023large} or through activation-level steering
derived from mechanistic analysis of where security-relevant
representations arise~\cite{sandoval2026surgical}. Others adapt the
model directly through fine-tuning, whether on curated
vulnerability-fixing examples~\cite{he2024instructiontuningsecurecode},
synthetic CWE-paired data~\cite{hajipour2024hexacoder}, or a small,
identified subset of security-relevant
neurons~\cite{thang2026goodvibe}. Across this line of work, generated
code is evaluated against known weakness patterns rather than by
asking the model to predict a CWE label as an explicit, reusable
output. HexaCoder is closest in spirit to our repair agent, since it
also pairs a specific CWE with a targeted code change, but does so
as training data baked into the model's weights rather than as an
inference-time diagnosis.

Producing a CWE label is fundamentally a closed-set decision, yet
the dominant way to obtain one from a language model is to generate
free-form text and parse it into a label, a mismatch discussed
further in Section~\ref{sec:jev-classification}. Jev's typed
\texttt{Choice} interface addresses this mismatch by evaluating a
declared candidate set directly and returning a probability for
each candidate, rather than requiring text generation followed by
extraction~\cite{almeida2026jev}. To our knowledge, no prior work
evaluates a typed decision model of this kind against autoregressive
language models specifically on CWE classification or on a
downstream, diagnosis-guided code-repair task; this comparison is
the focus of the present paper.
\section{Discussion }
\label{discussion}
\textbf{The best classifier depends on how its predictions are used.}
GPT-5.6-Sol achieves the highest Top-1 accuracy (0.495) and Macro-F1 (0.360), whereas Jev performs best on Top-3 (0.766) and Top-5 accuracy (0.836). Their MRR scores are nearly identical (0.628 and 0.624, respectively), indicating similar overall ranking quality. These results suggest that GPT-5.6-Sol is more accurate when only exact Top-1 CWE classification is considered, while Jev more frequently places the correct CWE among its highest-ranked alternatives. Importantly, however, Top-1 classification accuracy does not directly determine downstream repair effectiveness. Despite its lower Top-1 accuracy, Jev-guided repair achieves a higher overall security pass rate than LLM-guided repair (0.707 vs. 0.661). This suggests that a prediction need not exactly match the reference CWE to provide useful repair guidance. Jev's \texttt{Choice} interface additionally provides probabilities over all 50 CWE classes, although our results do not establish whether this interface itself explains the observed differences.

\textbf{Both approaches can provide predictive uncertainty, but at different computational costs.} In preliminary experiments, we derived a normalized distribution over the 50 candidate CWEs from autoregressive models by scoring each candidate separately and applying a softmax to their sequence log-likelihoods. This enables uncertainty and calibration analysis, but was substantially more expensive than the single-generation ranking used in our main experiments, making it impractical for our online repair pipeline. Jev instead returns scores over the declared alternatives directly through its typed \texttt{Choice} interface within a single request~\cite{almeida2026jev}.

\textbf{Efficiency comes with a narrower scope.}
In our main experiments, Jev reduces median API latency by $6.27\times$ and estimated API cost by $55.9\times$ relative to GPT-5.6-Sol. These measurements characterize the specific systems and API configurations evaluated here and should not be interpreted as an architecture-independent efficiency advantage. They compare Jev with our single-generation autoregressive setup rather than the preliminary procedure that separately scores all 50 candidates. Jev also requires the candidate set and output type to be specified in advance and does not support open-ended text generation. It is therefore best viewed as a specialized component for fixed-choice decisions rather than a general replacement for autoregressive language models.

\textbf{Probability outputs do not establish calibration.}
Although Jev provides a distribution over the complete candidate set and its developers describe it as trained for calibrated decisions, our experiments do not establish whether its reported confidence consistently corresponds to empirical correctness. Evaluating this claim would require a dedicated calibration study across multiple decision domains.

\section{Conclusion }
\label{conclusion}
We evaluated a typed decision model, Jev, against a range of
autoregressive language models on a controlled 50-way CWE
classification task, and examined whether differences in
classification quality translate into measurable improvements when
used to guide automatic code repair. Jev substantially outperforms
every evaluated open-weight autoregressive model across all
classification and ranking metrics, and is competitive with a
frontier proprietary model, GPT-5.6-Sol: GPT-5.6-Sol achieves higher
Top-1 accuracy and Macro-F1, while Jev achieves higher Top-3 and
Top-5 accuracy and a nearly identical MRR, at approximately
$55.9\times$ lower estimated API cost and $6.27\times$ lower median
latency. When used to guide a diagnosis-and-repair agent, replacing an
autoregressive CWE classifier with Jev increases the overall
security pass rate from 0.661 to 0.707 and fixes 228 initially
vulnerable programs, compared with 129 under LLM-guided repair.
However, both approaches occasionally introduce new vulnerabilities,
and neither improves security consistently across all languages.

In our experiments, Jev provides faster and cheaper CWE classification than all tested open-weight models and GPT-5.6-Sol, while returning valid predictions for every example. Classification performance varies by metric: GPT-5.6-Sol achieves the highest Top-1 accuracy and Macro-F1, while Jev achieves the highest Top-3 and Top-5 accuracy, with similar MRR. These findings suggest that Jev is useful for automated tasks requiring quick, structured choices from a fixed set, rather than as a general replacement for autoregressive models.

Our conclusions are subject to several limitations. The downstream
evaluation uses a single repair model and only the Top-1 diagnosis
from each classifier, so results may not generalize to other repair
backbones or to repair guided by a full ranked list. Both our
classification and downstream results rely on a single benchmark,
CyberSecEval's Instruct subset, and a single external evaluator,
whose CWE and insecurity judgments are imperfect proxies for
manually verified ground truth. Jev is a commercial system
documented only by its vendor, with no independent peer-reviewed
evaluation of its architecture or training; while it is described as
trained for calibrated decisions, our experiments do not establish
general calibration. Finally, our cost and latency figures reflect
one provider's pricing at the time of experimentation and may not
hold under different pricing or serving conditions. Future work
should evaluate additional repair backbones, exploit the
hierarchical structure of the CWE taxonomy, conduct a dedicated
calibration study, and seek independent evaluation of Jev and
similar typed decision models across further benchmarks

\clearpage
\bibliographystyle{IEEEtran}
\bibliography{bibliography}

\end{document}